\documentclass[10pt,nofootinbib]{revtex4-2}
\usepackage{graphicx} 
\usepackage{color}
\usepackage{babel}
\usepackage{fancybox}
\usepackage{calc}
\usepackage{amsmath}
\usepackage{amssymb}
\usepackage[a4paper]{geometry}
\usepackage{comment}
\usepackage[unicode=true,pdfusetitle,
 bookmarks=true,bookmarksnumbered=false,bookmarksopen=false,
 breaklinks=true,pdfborder={0 0 0},pdfborderstyle={},backref=false,colorlinks=true]
 {hyperref}

\begin{document}

\title{\Large 
Time-dependent Trapped Plasmas: Nonlinear Dynamics, Symmetries and Invariants
}

\author{Thonimar V. Alencar}
\affiliation{Grupo de Física Teórica e Computacional, Departamento de Ciências Naturais, CEUNES, Universidade Federal do Espírito Santo (UFES), Rodovia Governador Mário Covas, Km 60, São Mateus, 29932-540, ES, Brasil}

\author{Luiz Gustavo Ferreira Soares}
\affiliation{Grupo de Física Teórica e Computacional, Departamento de Ciências Naturais, CEUNES, Universidade Federal do Espírito Santo (UFES), Rodovia Governador Mário Covas, Km 60, São Mateus, 29932-540, ES, Brasil}

\author{Ronaldo Thibes}
\affiliation{Departamento de Ci\^encias Exatas e Naturais,
Universidade Estadual do Sudoeste da Bahia,
 45700-000, Itapetinga BA, Brazil}

\begin{abstract}

We investigate the nonlinear dynamics of a single-component plasma confined in a time-dependent harmonic trap regarding aspects of symmetry and invariant functions. The system is described as a fluid in an isentropic adiabatic regime by a system of partial differential equations.  A convenient change of variables, with a Gaussian ansatz for the number density distribution, allows a consistent mathematical description in terms of ordinary
differential equations, from which we follow up with an analysis  concerning the corresponding differential operators algebraic structure and Noether symmetries in specific physical regimes. For each studied case, proper invariants
are identified.  The obtained conserved quantities capture
an interplay between the internal plasma dynamics and the
time modulation of the trap, 
 resulting in
a sharp restriction for the system evolution. 

\end{abstract}

\maketitle

~

\section{Introduction}

Nonlinear dynamical systems have been spreading throughout physics and science in general for more than a couple of centuries.
Their mathematical structures and symmetries still pose countless quest challenges in present-day mathematics, physics and mathematical-physics \cite{MarsdenRatiu1999, Euler, EulerNucci, EulerZhang, 
Rajeev:2025dos, Henneaux:1992ig, tHooft:1994pio, Berghofer:2021ufy, Sorella:2010it, Reshetnyak:2013bga, Mandal:2023pdk, Mandal:2025vfs, Saha:2008iz, Chernyavsky:2019hyp, Belich:2025qbu, Batlle:2018hoe, Reis:2026rch, CesareSilva:2026zba, Shapiro:2025pmn, Pasti:1995tn, Krechet:2015ara}.
In particular, in modern times, exact symmetries and controlled symmetry violations have been playing key roles concerning both their proper fundamental structures as well as specific consequences and applications.  For that matter, we may mention instances of gauge \cite{Henneaux:1992ig, tHooft:1994pio, Berghofer:2021ufy}, quantum
\cite{Sorella:2010it, Reshetnyak:2013bga, Mandal:2023pdk, Mandal:2025vfs}, Galilean \cite{Saha:2008iz, Chernyavsky:2019hyp, Belich:2025qbu, Batlle:2018hoe}, Lorentz \cite{Batlle:2018hoe, Reis:2026rch, CesareSilva:2026zba}, conformal/dual \cite{Shapiro:2025pmn, Pasti:1995tn, Krechet:2015ara} and many others perfectly realized or broken controlled symmetries in dynamical systems throughout contemporary physics.  Of special importance for the present work, Lie and Noether symmetries \cite{LeachPaliathanasis, olver1993applications, BlumanAnco, Cantwell} play a distinct role, allowing for the characterization and classification of dynamical systems and leading to conserved or invariant quantities.
Coming closer to our point, the use of symmetry methods in the physics of plasmas, Bose-Einstein condensates and related systems has produced important recent breakthrough results \cite{Haas1, Qin:2006an, Haas2, Soares:2026wrk}, promoting an approximation between distinct fields of mathematics and physics, while inviting new interdisciplinary collaborative works. Besides their intrinsic beauty, the identification of conserved quantities in such systems has practical importance, which can lead to technological applications.

Building on and extending previous works \cite{Soares:2026wrk, Soares:2020, Soares:2021pfm}, we consider a quasi-one-dimensional {single-component} plasma, trapped by a time-dependent confining potential, described by a system of partial differential equations (pde).  The problem can be formulated in a variational Lagrangian framework allowing the consideration of a simplifying realistic ansatz for the particle number density, after which the original pde system decouples into a pair of ordinary differential equations concerning the center-of-mass and spreading width modes.  Turning to phase space, we reduce the order of the evolution equations, obtain the corresponding differential operator vector fields, and proceed to their algebraic analysis.  By splitting the effective Lagrangian into two independent parts and considering cases of physical interest, namely thermal and cold plasmas, we obtain three invariant functions. 
The latter correspond to  conserved quantities configuring restrictions on plasma dynamical evolution through the confining trap's time-modulation.

After this brief Introduction, we start our technical discussion in the remaining article with a presentation organized as follows. In Sec.~II, we present the plasma model as described by a system of three coupled pdes plus a constitutive state equation and show how a convenient Gaussian ansatz for the particle number density leads to simpler ordinary differential equations.  In Sec~III, after the introduction of proper auxiliary variables, the differential operators commutator algebra associated with the first order odes is investigated. 
The search for symmetries and conserved quantities is reserved for Sec.~IV, in which we split the Lagrangian into two independent parts corresponding to the plasma center of mass and spreading width.  Still in Sec. IV, by applying Noether's invariance condition, we find the necessary odes to produce invariant quantities associated to
the remaining degrees of freedom in relevant physical regimes.  We end in Sec.~V with our final remarks and conclusion.

\section{The Plasma Model}
The time $t$ evolution of a one-dimensional single-component plasma, composed of constituent particles of mass $m$ and charge $e$ moving along a linear direction $z$, can be described in terms of its linear corpuscle number density $n(z,t)$, velocity field $v(z,t)$, and self-consistent potential $\phi(z,t)$. We consider the model recently discussed in \cite{Soares:2026wrk}, in which those three real functions are interrelated through the pde system
\begin{eqnarray}\label{1.a}
\frac{\partial n}{\partial t} + \frac{\partial}{\partial z}(nv) &=& 0\,, \\
\label{1.b}
\frac{\partial v}{\partial t} + v\frac{\partial v}{\partial z} + \frac{\Sigma_\perp}{m n}\frac{\partial p}{\partial z} {+}  \frac{e}{ m}\frac{\partial \phi}{\partial z}&=&  -\frac{1}{m}\frac{\partial }{\partial z}V_c(z,t)\,,\\
\label{1.c}
\frac{\partial^2 \phi }{\partial z^2} {+} \frac{e n}{\varepsilon_0 \Sigma_\perp}&=&0 \,,
\end{eqnarray}
complemented by the isentropic equation of state
\begin{equation}\label{ES}
p \Sigma_\perp = n_0 k_B T_0 \left(\frac{n}{n_0}\right)^3\,,
\end{equation}
in which $\epsilon_0$ and $k_B$ represent physical universal constants, 
$n_0$ and $T_0$ fixed reference values, $\Sigma_\perp$ denotes the surface area perpendicular to the $z$ direction and $V_c(z,t)$ stands for a given control 
 potential function.  The last four equations form a closed system.  For an adiabatic cooling regime, common in many relevant realistic models \cite{Gabrielse:2011qgf, manfredi2012adiabatic},
we consider a slow-time-dependent harmonic potential
\begin{equation}\label{1.c0}
V_c =\frac{m\omega^2(t)z^2}{2}\,,
\end{equation}
with a time-decreasing frequency
\begin{equation}\label{omega(t)}
\omega(t)=\frac{\omega_0}{(1 + \Omega t)^{\beta}} \,,
\end{equation}
for positive constants $\omega_0$, $\Omega$ and $\beta \leq 1$. We also require $\beta\Omega \ll \omega_0$, leading to  $|\dot\omega| \ll \omega^2$, 
ensuring a slowly varying energy regime. A proper discussion of the physics involved in (\ref{1.a})-(\ref{1.c}) can be found in
\cite{Soares:2026wrk, Soares:2020}.

As a mathematical model, the pde system (\ref{1.a})-(\ref{1.c}) can be obtained from a variational principle.  In fact, we may define an action functional
\begin{eqnarray} \label{S}
{\cal S}[n, \phi, \theta] &=&  \int dt\,dz\,
\bigg\{
\frac{mn}{2}\bigg(\frac{\partial \theta}{\partial z} \bigg)^2+mn\frac{\partial \theta}{\partial t}+n(V_c{+}e\phi) \nonumber\\
&&
-\frac{\epsilon_0 \Sigma_\perp}{2}\bigg(\frac{\partial \phi}{\partial z}\bigg)^2 + \Sigma_\perp \int dn \int \frac{d p}{n}
\bigg\}\,,
\end{eqnarray}
in terms of the three two-variable real functions
\begin{equation}\label{fields}
 n(z,t), ~~~~ \phi(z,t), ~~~~ \theta(z,t)\,,
\end{equation}
with $n(z,t)$ and $\phi(z,t)$ as before, and $\theta(z,t)$ satisfying ${v}=\partial \theta/\partial z$, demanding its stationarity with respect to first order functional variations in \eqref{fields}.  Then, taking (\ref{ES}) into account, the corresponding Euler-Lagrange field equations lead to the system of partial  differential equations (\ref{1.a})-(\ref{1.c}).

In most situations of physical interest, the plasma linear density follows a Gaussian distribution along the $z$-direction, with time-dependent mean and variance.  Hence, we consider a collective behavior 
\begin{eqnarray}\label{1.e}
    n(z,t)=\frac{N}{\sqrt{2\pi}\alpha(t)}\exp\bigg({-\frac{(z-d(t))^2}{2\alpha^2(t)}}\bigg)\,,
\end{eqnarray}
leading to a dynamical description in terms of the two degrees of freedom $d(t)$ and $\alpha(t)$, characterizing the plasma distribution center and spreading as functions of time.
The normalization constant
$N$ in \eqref{1.e}
represents the total number of constituent units for the confined plasma.  Once \eqref{1.e} is given, on account of the original differential equations system (\ref{1.a})-(\ref{1.c}), the plasma dynamics is characterized through the remaining two-variable real functions as
\begin{eqnarray}
   \phi(z,t)&=&{-}\frac{ e N }{\epsilon_0\Sigma_\perp}\bigg[\frac{\alpha(t)}{\sqrt{2\pi}} \exp\bigg({-\frac{(z-d(t))^2}{2\alpha^2(t)}}\bigg) \nonumber\\
   &+&\frac{z-d(t)}{2}{ \mbox{erf}}\bigg({\frac{z-d(t)}{\sqrt{2}\alpha(t)}}\bigg)\bigg]\,, \label{1.h}
\end{eqnarray}
\begin{equation}\label{1.g}
\theta(z,t) = \frac{\dot{\alpha}(t)}{2\alpha(t)}(z-d(t))^2+\dot{d}(z-d(t))\,,
\end{equation}
and
\begin{equation}\label{1.f}
    v(z,t) = \frac{\dot{\alpha}(t)}{\alpha(t)}(z-d(t))+\dot{d} \,.
\end{equation} 
In this manner, the original problem of solving the pde system (\ref{1.a}-\ref{1.c}) has been conveyed to finding the two time-dependent functions $d(t)$ and $\alpha(t)$.

In order to derive the dynamical behavior of the time-dependent coordinates, we may integrate out the $z$ space-dependence in the action functional (\ref{S}) to obtain an effective Lagrangian per unit mass and particle number, defined as 
\begin{equation}\label{1.l0a}
    L (d,\alpha,\dot{d},\dot{\alpha})\equiv
    -\frac{1}{m N }\int\mathcal{L} dz
    \,,
\end{equation}
with $\cal L$ characterized by the integrand of \eqref{S}.  In this fashion, using
equations (\ref{1.e}), (\ref{1.h}) and (\ref{1.g}), and performing the corresponding $z$-integration in (\ref{1.l0a}), we obtain 
\begin{equation}\label{1.l0}
    L (d,\alpha,\dot{d},\dot{\alpha})=\frac{1}{2}(\dot{d}^2+\dot{\alpha}^2)-U_d-U_\alpha\,,
\end{equation}
with the two time-dependent effective potentials $U_d$ and $U_\alpha$ written as 
\begin{equation}\label{Ud}
    U_d\equiv \frac{\omega^2(t)}{2}d^2
\end{equation}
and
\begin{equation}\label{1.l}
    U_\alpha\equiv\frac{\omega^2(t)}{2}\alpha^2+\frac{a}{2\alpha^2}-b\alpha\,,
\end{equation}
in terms of the parameters
\begin{equation}\label{a}
a \equiv {k_BT_0 N^2}/({2\sqrt{3}\pi mn_0^2})
\end{equation}
and
\begin{equation}\label{b}
b\equiv {Ne^2}/(2\sqrt{\pi}\epsilon_0m\Sigma_\perp)\,.
\end{equation}
The effective potentials above $U_d$ and $U_\alpha$ correspond, respectively, to the system's dipole center-of-mass and spreading width oscillating modes, with the constants (\ref{a}) and (\ref{b}) characterizing thermal and electric effects.  The external time-dependent frequency $\omega(t)$ present in equations (\ref{Ud}) and (\ref{1.l}) controls the harmonic confinement as given by (\ref{omega(t)}).

Associated with \eqref{1.l0}, we have the two non-autonomous ordinary differential equations
\begin{equation}\label{d}
    \ddot d+\omega(t) d=0
\end{equation}
and
\begin{equation}\label{alpha}
    \ddot{\alpha} + \omega^2(t) \alpha - \frac{a}{\alpha^3} = b\,.
\end{equation}
The first one, equation \eqref{d}, corresponds to a time-dependent harmonic oscillator and has been extensively investigated in the literature
\cite{HRLewis:1967gqm, Lewis1968, lutzky1978noether, Leach1977, Leach:1980fa, Bertin2012, Bertin:2020mbb}, while \eqref{alpha} can be understood as a generalized or forced Pinney equation \cite{pinney1950nonlinear} due to the presence of the constant term $b$ in its right hand side.

We are interested in the subjacent plasma dynamics described by the odes \eqref{d} and \eqref{alpha},
whose analytical behavior, differential operator structures, symmetries, and invariants we discuss and contextualize in the remaining sections.

\section{Differential Operators Algebra}
In this section, we study the differential operator commutator structure associated with the plasma dynamics described by odes \eqref{d} and \eqref{alpha}.  As is well-known, given a system of second-order ordinary differential equations, by introducing auxiliary intermediate variables, it can be reduced to first-order, being written as
\begin{equation}\label{foodes}
    \frac{d x^r}{dt}=y^r(x,t)\,,\quad x=(x^r)\,,\quad \quad r=1,\dots,n\,,
\end{equation}
for specific functions $y^r(x,t)$ and a fixed $n\in \mathbb{N}$.  Once cast in the form \eqref{foodes}, we can associate it with the vector field
\begin{equation}
    \mathbf{X}=\sum_{r=1}^n y^r(x,t)\frac{\partial}{\partial x^r}\,.
\end{equation}
We say that $\mathbf{X}$ admits a superposition rule in terms of a set of $m$ linearly independent (li) operators $X_i$, $i=1,\dots,m$, with $m\in\mathbb{N}$, when it can be locally written as
\begin{equation}\label{X}
    \mathbf{X}(x,t)=\sum_{i=1}^mp^i(t)X_i(x)\,,
\end{equation}
for some $p^i(t)$.
The differential operators $X_i$, $i=1,\dots,m$, possibly li completed with $X_j$, $j=m+1,\dots,d$, $d\geq m$, generate a $d$-dimensional Lie algebra $\Lambda^d$ if there exist $c_{ij}^{\,\,\,\,k}$, $i,j,k=1,\dots,d$, such that
\begin{equation}\label{LA}
    [X_i, X_j]=\sum_{k=1}^dc_{ij}^{\,\,\,\,k} X_k
    \,,
\end{equation}
with the square bracket denoting the usual commutator, i.e.,
\begin{equation}
    [X_i, X_j] \equiv X_i\,X_j - X_j\,X_i
    \,.
\end{equation}
In the following, we investigate the vector field $\mathbf{X}$ and the corresponding superposition rule structure for specific cases of \eqref{d} and \eqref{alpha} of physical interest.
\subsection{Center of Mass}
The plasma center of mass motion is described by the non-autonomous linear homogeneous ordinary differential equation (ode) \eqref{d}, with an external time-dependent frequency $\omega(t)$ given by \eqref{omega(t)}. Introducing the auxiliary variable $v$, by demanding $v=\dot{d}$, the second-order ode \eqref{d} can be rewritten as
   \begin{equation}\label{l1}
        \begin{cases}
    \dot d=v  \,,   \\
    \dot v=-\omega(t)^2d  \,,
  \end{cases}
   \end{equation}
with the associated vector field
\begin{equation}
     \mathbf{X}(d,v,t)=v\frac{\partial}{\partial d}-\omega(t)^2d\frac{\partial}{\partial v}
     \,.
\end{equation}
This is clearly in the form \eqref{X}
with $p_1=-\omega^2(t)$, $p_2=1$, and 
\begin{equation}\label{X1X2}
    X_1\equiv d\frac{\partial}{\partial v}\,, \,\,\,\,X_2\equiv v\frac{\partial}{\partial d} \,.
\end{equation}
The two operators in \eqref{X1X2} above satisfy the commutation relation
\begin{equation}
    [X_1,X_2]= d\frac{\partial}{\partial v}\bigg(v\frac{\partial}{\partial d}\bigg)-v\frac{\partial}{\partial d}\bigg(d\frac{\partial}{\partial v}\bigg)= X_3\,,
\end{equation}
where we have defined
\begin{equation}
    X_3
    \equiv d\frac{\partial}{\partial d}-v\frac{\partial}{\partial v} \,.
\end{equation}
Furthermore, we have
\begin{equation}
      [X_1,X_3]=d\frac{\partial}{\partial v}\bigg[\bigg(d\frac{\partial}{\partial d}-v\frac{\partial}{\partial v}\bigg)\bigg]-\bigg(d\frac{\partial}{\partial d}-v\frac{\partial}{\partial v}\bigg)\bigg(d\frac{\partial}{\partial v}\bigg)=-2X_1\,,
\end{equation}
and
\begin{equation}
    [X_2,X_3]=v\frac{\partial}{\partial d}\bigg[\bigg(d\frac{\partial}{\partial d}-v\frac{\partial}{\partial v}\bigg)\bigg]-\bigg(d\frac{\partial}{\partial d}-v\frac{\partial}{\partial v}\bigg)\bigg(v\frac{\partial}{\partial d}\bigg)=2X_2\,.
\end{equation}
From these results, we see that the first-order system \eqref{l1} closes a three dimensional Lie algebra \eqref{LA} with non-null structure constants 
\begin{equation}\label{cs}
c_{12}^{\,\,\,\,\,\,3}=1\,,\quad c_{13}^{\,\,\,\,\,\,1}=-2\,,\quad   c_{23}^{\,\,\,\,\,\,2}=2\,.
\end{equation}

\subsection{Spreading Width}
Concerning the plasma spreading width $\alpha(t)$ introduced in \eqref{1.e} as a standard deviation for the particle density distribution, we split our analysis into the particular cases of thermal and cold plasmas, respectively described by $b=0$ and $a=0$. 
\subsubsection{{Thermal Plasma}}   
In this case, as we have $b=0$ and $a \neq 0$, the second-order ode \eqref{alpha} is non-linear and can be rewritten in first order as
   \begin{equation}\label{l2}
        \begin{cases}
    \dot \alpha=v  \,,   \\
    \displaystyle \dot v=-\omega(t)^2\alpha+\frac{a}{\alpha^3} \,,
  \end{cases}
   \end{equation}
   with $v$ representing an auxiliary variable.  We associate to \eqref{l2} the vector field
\begin{equation}
     \mathbf{N}=v\frac{\partial}{\partial \alpha}-\bigg(\omega(t)^2\alpha-\frac{a}{\alpha^3}\bigg)\frac{\partial}{\partial v}
     \,,
\end{equation}
which is of the form \eqref{X}
with $p_1=-\omega^2(t)$, $p_2=1$, and
\begin{equation}
    N_1\equiv\alpha\frac{\partial}{\partial v}\,,\,\,\,  N_2 \equiv \frac{a}{\alpha^3}\frac{\partial}{\partial v}+v\frac{\partial}{\partial \alpha}\,.
\end{equation}
For the commutation relations, defining
\begin{equation}
N_3 \equiv \alpha\frac{\partial}{\partial \alpha}-v\frac{\partial}{\partial v}
    \,,
\end{equation}
we have
\begin{equation}
    [N_1,N_2]= \alpha\frac{\partial}{\partial v}\bigg(\frac{a}{\alpha^3}\frac{\partial}{\partial v}+v\frac{\partial}{\partial \alpha}\bigg)-\bigg(\frac{a}{\alpha^3}\frac{\partial}{\partial v}+v\frac{\partial}{\partial \alpha}\bigg)\alpha\frac{\partial}{\partial v}= N_3\,,
\end{equation}
\begin{equation}
      [N_1,N_3]=\alpha\frac{\partial}{\partial v}\bigg[\bigg(\alpha\frac{\partial}{\partial \alpha}-v\frac{\partial}{\partial v}\bigg)\bigg]-\bigg(\alpha\frac{\partial}{\partial \alpha}-v\frac{\partial}{\partial v}\bigg)\bigg(\alpha\frac{\partial}{\partial v}\bigg)=-2N_1\,,
\end{equation}
and
\begin{equation}
    [N_2,N_3]=\bigg(\frac{a}{\alpha^3}\frac{\partial}{\partial v}+v\frac{\partial}{\partial \alpha}\bigg)\bigg(\alpha\frac{\partial}{\partial \alpha}-v\frac{\partial}{\partial v}\bigg)-\bigg(\alpha\frac{\partial}{\partial \alpha}-v\frac{\partial}{\partial v}\bigg)\bigg(\frac{a}{\alpha^3}\frac{\partial}{\partial v}+v\frac{\partial}{\partial \alpha}\bigg)=2N_2\,.
\end{equation}
Thus, the commutation relations above for the non-autonomous non-linear system \eqref{l2} also characterize a Lie algebra with structure constants \eqref{cs}.

\subsubsection{Cold Plasma}
In this second case, considering $a=0$ and $b\neq 0$, we may rewrite \eqref{alpha} in first order as
\begin{equation}\label{l3}
        \begin{cases}
    \dot \alpha=v     \\
    \dot v=-\omega(t)^2\alpha+b
  \end{cases}
   \end{equation}
with an associated vector field
\begin{equation}\label{M}
     \mathbf{M}=v\frac{\partial}{\partial \alpha}-(\omega(t)^2\alpha-b)\frac{\partial}{\partial v} \,. 
\end{equation}
Again, $v$ is an auxiliary variable connecting the two first-order equations \eqref{l3} that ensure equivalence to \eqref{alpha}.
Defining
\begin{eqnarray}
     &&M_1=\alpha\frac{\partial}{\partial v}\,,\,\,\,  M_2=v\frac{\partial}{\partial x}\,,\,\,\,\,\,M_4=b\frac{\partial}{\partial v}\,.
\end{eqnarray}
we may rewrite \eqref{M} as
\begin{equation}
     \mathbf{M}=M_2-\omega(t)^2M_1+M_4\,,
\end{equation}
which is in the form \eqref{X} with $p_1=-\omega^2(t)$, $p_2=p_4=1$, evincing a superposition rule admission.
By defining further
\begin{equation}
    M_3 \equiv \alpha\frac{\partial}{\partial \alpha}-v\frac{\partial}{\partial v}\,,
\end{equation}
\begin{equation}
 M_5 \equiv    b\frac{\partial}{\partial \alpha} \,,
\end{equation}
it is straightforward to obtain the non-null commutation relations
\begin{equation}
    [M_1,M_2]= \alpha\frac{\partial}{\partial v}\bigg(v\frac{\partial}{\partial \alpha}\bigg)-v\frac{\partial}{\partial \alpha}\bigg(\alpha\frac{\partial}{\partial v}\bigg)=  M_3\,,
\end{equation}
\begin{equation}
      [M_1,M_3]=\alpha\frac{\partial}{\partial v}\bigg[\bigg(\alpha\frac{\partial}{\partial \alpha}-v\frac{\partial}{\partial v}\bigg)\bigg]-\bigg(\alpha\frac{\partial}{\partial \alpha}-v\frac{\partial}{\partial v}\bigg)\bigg(\alpha\frac{\partial}{\partial v}\bigg)=-2M_1\,,
\end{equation}
\begin{equation}
    [M_2,M_3]=v\frac{\partial}{\partial \alpha}\bigg[\bigg(\alpha\frac{\partial}{\partial \alpha}-v\frac{\partial}{\partial v}\bigg)\bigg]-\bigg(\alpha\frac{\partial}{\partial \alpha}-v\frac{\partial}{\partial v}\bigg)\bigg(v\frac{\partial}{\partial \alpha}\bigg)=2M_2\,.
\end{equation}
\begin{equation}
    [M_4,M_2]=b\frac{\partial}{\partial v}\bigg(v\frac{\partial}{\partial \alpha}\bigg)-v\frac{\partial}{\partial \alpha}\bigg(b\frac{\partial}{\partial v}\bigg)=b\frac{\partial}{\partial \alpha}=M_5\,,
\end{equation}
\begin{equation}
    [M_4,M_3]=b\frac{\partial}{\partial v}\bigg(\alpha\frac{\partial}{\partial \alpha}-v\frac{\partial}{\partial v}\bigg)-\bigg(\alpha \frac{\partial}{\partial \alpha}-v\frac{\partial}{\partial v}\bigg)\bigg(b\frac{\partial}{\partial v}\bigg)=-{M_4}\,.
\end{equation}
\begin{equation}
      [M_5,M_1]=b\frac{\partial}{\partial \alpha}\bigg(\alpha\frac{\partial}{\partial v}\bigg)-\alpha\frac{\partial}{\partial v}\bigg(b\frac{\partial}{\partial \alpha}\bigg)=M_4\,,
\end{equation}
and
\begin{equation}
    [M_5,M_3]=b\frac{\partial}{\partial \alpha}\bigg(\alpha\frac{\partial}{\partial \alpha}-v\frac{\partial}{\partial v}\bigg)-\bigg(\alpha\frac{\partial}{\partial \alpha}-v\frac{\partial}{\partial v}\bigg)\bigg(b\frac{\partial}{\partial \alpha}\bigg)={M_5}\,.
\end{equation}
Hence, the previous results can be summarized in the form \eqref{LA} with
\begin{equation}
\begin{gathered}
c_{12}^{\,\,\,\,\,\,3}=1 \,,\,\,\,
c_{13}^{\,\,\,\,\,\,1}=-2\,,\,\,\,
c_{15}^{\,\,\,\,\,\,4}=-1\,,\,\,\,
c_{23}^{\,\,\,\,\,\,2}=2\,,\,\,\,
\\
c_{24}^{\,\,\,\,\,\,5}=-1\,,\,\,\,
c_{34}^{\,\,\,\,\,\,4}=1\,,\,\,\,
c_{35}^{\,\,\,\,\,\,5}=-1\,,\,\,\,
\end{gathered}
\end{equation}
showing that, in the cold plasma case, the differential operators close a $5-$dimensional Lie algebra. 

\section{NOETHER INVARIANTS}
Since the two relevant variables $d$ and $\alpha$ are not coupled to each other in \eqref{1.l0}, it is possible to split the Lagrangian function  as
\begin{equation}\label{L12}
    L (d,\alpha,\dot{d},\dot{\alpha})= L_1(d,\dot{d}) + L_2(\alpha,\dot{\alpha})\,,
\end{equation}
with
\begin{equation}
     L_1(d,\dot{d}) = \frac{\dot{d}^2}{2}-U_d\quad\mbox{and}\quad
    L_2(\alpha,\dot{\alpha}) = \frac{\dot{\alpha}^2}{2}-U_\alpha\,,
\end{equation}
and, accordingly, look for conserved quantities associated with the symmetries of each separate constituent part.
As Noether has taught us in her beautiful theorems \cite{Noether, Schwarzbach2011, Halder:2018kyp}, invariance of the action under continuous transformations leads to conserved quantities.
Along that line, following a well-established standard route \cite{Haas2, lutzky1978noether, ray1979noether, ray1982invariants, SarletCantrijn}, we consider $\epsilon$-parametrized transformations of the form
\begin{equation}\label{IT}
\begin{cases}
    t ~~\longrightarrow~~ t'=t+\epsilon T(x,t)\,,
\\
x ~~ \longrightarrow~~ x'=x+\epsilon\eta(x,t)\,,
\end{cases}
\end{equation}
with $x=d\,,\alpha$, and demand invariance of the action functional
\begin{equation}\label{S31}
S = \int d t L_{} \,,
\end{equation}
with $L=L_1\,,L_2$.
Associated with\eqref{IT}, we define the zero-order point symmetry generator 
\begin{equation}\label{G}
    G^{[0]}=T\frac{\partial}{\partial t}+\eta\frac{\partial}{\partial x}
\end{equation}
and its corresponding  first-order prolongation \cite{olver1993applications, BlumanAnco}
\begin{equation}\label{G1}
   G^{[1]}= G^{[0]}+(\dot \eta - \dot T \dot{x})\frac{\partial}{\partial \dot{x}}\,.
\end{equation}
The requirement of invariance of (\ref{S31}) under \eqref{IT} leads to the existence of a function $F=F(x,t)$ satisfying the Noether condition
\begin{equation}\label{NC}
    G^{[1]}L+\dot T L=\frac{\partial F}{\partial t}+\dot{x}\frac{\partial F}{\partial x}
\end{equation}
and to the consequent time invariant (conserved quantity) combination
\begin{equation}\label{I}
     I=T\bigg(\dot{x} \frac{\partial L}{\partial \dot{x}}-L\bigg)-\eta\frac{\partial L}{\partial \dot{x}}+F\,.
\end{equation}

To proceed further,
similarly to the previous section, it is convenient to separate the symmetry analysis into specific cases of physical interest.

\subsection{Center of Mass}
Concerning the center of mass mode, we first restrict our attention to transformations which do not affect $\alpha$
and apply \eqref{IT} to $L_1$ with $x=d$. Imposing Noether's symmetry condition \eqref{NC}, we obtain
\begin{equation}
    T=T(t)\,,\quad
    \eta(d,t) =\frac{{\dot T}(t)d}{2}-g(t)\,,
\end{equation}
and
\begin{equation}\label{71}
    F(d,t)=\frac{\ddot{T} d^2}{4}-\dot gd\,,
\end{equation}
with $g(t)$ denoting a new time-dependent function that satisfies
\begin{equation}\label{g}
    \ddot{g}+{\omega}^2(t)g=0\,.
\end{equation}
As a further consequence of \eqref{NC}, $T$ must also satisfy
\begin{equation}\label{T}
    \dddot{T}+4 \omega^2\dot T+4{\dot \omega}\omega T=0\,.
\end{equation}
This last equation may be further simplified by means of a change of variables $T=\rho^2$ allowing a time integration and consequent order reduction in terms of an integration constant $k$, to
\begin{equation}\label{P}
    \ddot \rho+ \omega^2\rho=\frac{k}{\rho^3}\,,
\end{equation}
which is recognized as the famous Pinney\footnote{\hspace{-0.1cm}Equation \eqref{P} is also more properly known as the Ermakov-Milne-Pinney equation \cite{pinney1950nonlinear, Ermakov1880, Milne}.} nonlinear equation \cite{CLR, carinena, MorrisLeach}.
The complete symmetry group of (\ref{P}) has been worked out by Nucci and Leach on \cite{Nucci2005}.  

Eventually, from (\ref{I}) applied to $L_1$ with $x=d$,
we obtain an invariant quantity given by   
\begin{equation}
     I=\frac{1}{2}\bigg[T\dot d^2-\dot T\dot d d+(\ddot T +2\omega^2T)\frac{d^2}{2}\bigg]+g\dot d-\dot gd\,\,,
\end{equation}  
which, in terms of the auxiliary variable $\rho$ satisfying \eqref{P}, can be rewritten as 
\begin{equation}\label{I_d}
    I=\frac{1}{2}(\rho\dot d-\dot\rho d)^2+\frac{k}{2}\bigg(\frac{d}{\rho}\bigg)^2+ g\dot d-\dot gd\,.
\end{equation}
Since $g$ is any solution to \eqref{g}, we may take $g=0$ for simplicity and obtain a first conserved quantity associated with the center-of-mass mode $I_0$ given by
\begin{equation}
    I_0=\frac{1}{2}(\rho\dot d-\dot\rho d)^2+\frac{k}{2}\bigg(\frac{d}{\rho}\bigg)^2
    \,.
\end{equation}

\subsection{Plasma's spreading width}
To next discuss the plasma spreading width mode $\alpha(t)$, we apply the transformation \eqref{IT} with $x=\alpha$ to $L_2$, and the potential subdivides into the thermal and cold plasma cases. 
\subsubsection{Thermal Plasma}
With a null charge $e=0$ for the constituent particles, a thermal plasma is characterized by $b=0$ in the effective potential (\ref{1.l}).  For this physical realization, we look for $L_2$  corresponding symmetries  generated by \eqref{IT} with $x=\alpha$.
In this case, the Noether symmetry condition (\ref{G}) leads to
\begin{equation}
    T=T(t)\,,\quad\eta=\frac{\dot{ T} \alpha}{2}\,,\quad
    F=\frac{\ddot{T} \alpha^2}{4} \,,
\end{equation}
with $T$ satisfying the same previous third-order differential equation (\ref{T}).  

Hence, associated with the thermal plasma spreading width, using equation (\ref{I}), we obtain the conserved quantity
\begin{equation}\label{CQ1}
    I=\frac{T}{2}\left(\dot{\alpha}^2+\omega^2 {\alpha}^2 + a\alpha^{-2} \right)-\frac{\dot T\alpha \dot \alpha} {2}+\frac{\ddot T\alpha^2}{4}\,.
\end{equation}
With a change of variables $T=\rho^2$, after a time integration introducing a constant $k$, the invariant \eqref{CQ1} can be reshaped as
\begin{equation}
    I_1 = \frac{1}{2}(\rho\dot{\alpha}-\dot{\rho} \alpha)^2+\frac{k}{2}\bigg(\frac{\alpha}{\rho}\bigg)^2 + \frac{a}{2}\bigg(\frac{\rho}{\alpha}\bigg)^2
    \,,
\end{equation}
with $\rho$ standing for a solution of the Pinney equation \eqref{P} with a corresponding $k$.  By using the equations of motion, it can be explicitly checked that the above quantity is conserved along the time evolution.

\subsubsection{Cold Plasma}
Finally, we consider the cold plasma situation in which we have $k_B T_0 = 0$, resulting in $a=0$ in (\ref{1.l}).  In this case, Noether's symmetry condition for $L_2$ under (\ref{NC}) leads to
\begin{equation}
    T=T(t)\,,\quad\eta=\frac{\dot{ T} \alpha}{2}-h(t)\,,\quad
    F=\frac{\ddot{T} \alpha^2}{4}-\dot{h}\alpha +m(t)\,,
\end{equation}
with the time-dependent functions $h(t)$ and $m(t)$ satisfying the differential equations
\begin{equation}
     \ddot{h}+{\omega}^2h=-\frac{3 b \dot T}{{2}}\,,\quad
     \dot{m}+bh=0\,,
\end{equation}
and $T$ stands for a solution of \eqref{T}.
The corresponding invariant, obtained from (\ref{I}), reads
\begin{eqnarray}
    I&=&\frac{1}{2}\left(T\dot{\alpha}^2-\dot{ T}\dot{\alpha}\right)  \alpha+\frac{1}{4}\left(\ddot{T} +2\omega^2T\right){\alpha^2}\nonumber\\
    &&+h\dot\alpha -\dot h \alpha -b\alpha T + m\,\,.
\end{eqnarray}
Similar to the previous cases, the transformation of variables $T=\rho^2$ allows for a first integration in terms of an integration constant $k$ with $\rho$ satisfying the Pinney equation \eqref{P}.  In terms of $\rho$, we can rewrite the above invariant as
\begin{equation}
    I_{2}=   {\frac{1}{2}(\rho\dot{\alpha}-\alpha\dot{\rho})^2}+\frac{k}{2}\bigg(\frac{\alpha}{\rho}\bigg)^2-\dot{h}\alpha+h\dot{\alpha}
    -b\alpha\rho^2+m\,.
\end{equation}
Thus, we have found an invariant corresponding to the cold plasma which is conserved along the system time evolution.

\section{Conclusion}
The mathematical structure of differential equations describing dynamical systems, particularly related to symmetries and invariants, can be successfully applied to hydrodynamic and ther\-mo\-dy\-nam\-ic systems.  In this paper, we have seen an explicit example related to plasma dynamics.  Starting from a pde model interrelating the relevant field variables describing a trapped one dimensional plasma under adiabatic evolution, we have shown that under some simplifying assumptions, the dynamics can be characterized by the time evolution of two time-dependent functions related to the plasma center-of-mass and spreading width through the mean variance of a Gaussian distribution for the linear particle density.  This allowed us to study the symmetries of simpler Lagrangians leading to corresponding Noether invariants.  We have also analyzed the algebraic aspects of the corresponding differential operators in terms of Lie algebras.  This route could also have been used to investigate the associated invariants, leading to the same obtained invariants in an equivalent way.  With the Lagrangians at hand, we have chosen to follow Noether's more intuitive approach directly associated with the invariance of the action under a set of parametrized infinitesimal transformations.  Further analysis concerning the mathematical structure of the ode system, as well as the relaxation of some of the simplifying assumptions for the pde systems are currently under analysis.

\subsection*{Acknowledgements}
\noindent R.T. gratefully acknowledges Prof Maria Clara Nucci for her two invited talks given at VI Ciclo de Seminários de Física em Itapetinga, which motivated a systematic follow up  search for lost symmetries in Nature.

\end{document}